\documentclass[aps,showpacs,twocolumn]{revtex4-1}
\usepackage{epsfig}
\usepackage{graphicx,color,dcolumn}
\usepackage{epstopdf}
\usepackage{amsmath,amssymb}
\usepackage{multirow}
\usepackage{diagbox}
\usepackage{changes}
\usepackage{booktabs}
\usepackage{threeparttable}
\begin{document}
\title{ $X(2900)$ in a chiral quark model}

\author{Yue Tan}
\email[E-mail: ]{181001003@stu.njnu.edu.cn}
\author{Jialun Ping}
\email[E-mail: ]{jlping@njnu.edu.cn (corresponding author)}
\affiliation{Department of Physics and Jiangsu Key Laboratory for Numerical
Simulation of Large Scale Complex Systems, Nanjing Normal University, Nanjing 210023, People's Republic of China}

\begin{abstract}
Recently, the LHCb Collaboration reported their observation of the first two fully open-flavor tetraquark states named $X_0(2900)$ and $X_1(2900)$
with unknown parity. Inspired by the report, we consider all of possible four-quark candidates of $X(2900)$ including molecular structure and
diquark structure in a chiral quark model with the help of Gaussian expansion method. Two different structures coupling is also considered.
To identify the genuine resonances, real-scaling method (stabilization method) was employed. The results show that no candidate of $X(2900)$ is
founded in $00^+$ and $01^+$ $cs\bar{q}\bar{q}$ system below the threshold of $D^*\bar{K}^*$, while there are two states in the $P$-wave excited $cs\bar{q}\bar{q}$ system, $D_1\bar{K}$
and $D_J\bar{K}$, which could be candidates of $X(2900)$. In this way, we assign negative parity to $X_1(2900)$,
and $X_0(2900)$ maybe a resonance state above the threshold of $D^*\bar{K}^*$, more calculation is needed.
\end{abstract}


\maketitle

\section{Introduction} \label{introduction}
The study of exotic hadron states, not only unveils the property of QCD, but also challenges the existing theories and models. However, it's a
controversial issue weather the exotic states exist?  Most of the exotic states can be classified in the traditional quark model and the properties
of the exotic states can be explained in the framework of quark model with some improvements. For example, the well-known exotic state \cite{Belle-3872,CDF-3872,Babar-3872,D0-3872}, $X(3872)$, can be explained as traditional $c\bar{c}$ state with large component $D\bar{D}^*+D^*\bar{D}$
in our unquenched quark model \cite{3872-tan}.

Actually, lots of experiment collaborations have been searching for exotic states in these two decades. In 2016, D0 collaboration observed a narrow
structure, which is denoted as $X(5568)$, in the $B_{s}^{0}\pi^{\pm}$ invariant mass spectrum with $5.1\sigma$ significance \cite{D0-5568}.
Because of the $B_{s}^{0}\pi^{\pm}$ decay mode, $X(5568)$ was interpreted as $s\bar{d}u\bar{b}$ ($s\bar{u}d\bar{b}$) tetraquark state. However,
it is difficult to find the candidate of $X(5568)$ in various approaches if requiring the ordinary hadrons can be described well in the approaches~\cite{PLB760}.
In our chiral quark model calculation all the possible candidates of $X(5568)$ are scattering states \cite{5568-1}, while we predicted one shallow
binding state \cite{5568-huang}, $B\bar{K}$ with $6.2$~ GeV, in the $00^+$ $bs\bar{q}\bar{q}$ system. Indeed, other experiment collaborations
did't find the evidence of $X(5568)$ \cite{Lhcb-5568}. Recently, the LHCb Collaboration coincidentally reported their observation of the first two fully
open-flavor tetraquark states named $X_0(2900)$ and $X_1(2900)$ in the $cs\bar{q}\bar{q}$ system, whose statistical significance is more than
$5\sigma$ \cite{Lhcb-2900,lhcb-2900-2}. If these two states are confirmed by the other Collaborations in the future, the $X(2900)$ could be the first exotic state
with four different flavors, which cannot be quark-antiquark system.
\begin{eqnarray}
&& X_0(2900):  M =2866\pm 7 {\rm MeV}, ~~\Gamma = 57\pm 3 {\rm MeV}, \nonumber \\
&& X_1(2900):  M= 2904\pm 5,{\rm MeV}, ~~\Gamma = 110\pm 12 {\rm MeV}. \nonumber
\end{eqnarray}

With the report of $X(2900)$, a lot of possible candidates spring up to explain $X(2900)$ in different frameworks \cite{molecular=baners,molecular=Burns,molecular=genglisheng,molecular=genglisheng2,molecular=Hejun,molecular=Mutuk,molecular=oset,molecular=oset2,molecular=sundu,molecular=Xue,tetraquark=Agaev,tetraquark=Chen,tetraquark=Karliner,tetraquark=Lu,tetraquark=Zhang,tetraquark=zhuruilin}, and most of them can be divided into two kinds of categories, dimeson structure and diquark structure. Y.~Xue {\em et~al.} got a $0^+$ $D^*\bar{K^*}$
resonance which can be the candidate of $X_0(2900)$ in the quark delocalization color screening model\cite{molecular=Xue}, and by means of qBSE approach,
J.~He {\em et~al.} also get the same conclusion \cite{molecular=Hejun}. M.~Karliner {\em et~al.} estimated roughly the diquark structure of $cs\bar{q}\bar{q}$,
and got a possible resonance which can be assigned as $X(2900)$ \cite{tetraquark=Karliner}, he also predicted a $0^+$ $bs\bar{q}\bar{q}$ resonance
with $6.2$~GeV. In the framework of QCD sum rule, H.~X.~Chen {\em et~al.} assigned $X_0(2900)$ as a $0^{+}$ $D^*\bar{K^*}$ molecular state while
$X_1(2900)$ as a $1^{-}$ $cs\bar{q}\bar{q}$ diquark state \cite{tetraquark=Chen}. However, with similar method, J.~R.~Zhang regarded both $X_0(2900)$ and
$X_1(2900)$ as diquark states \cite{tetraquark=Zhang}. In addition, before the report of $X(2900)$, S.~S.~Agaev {\em et~al.} obtained a unstable resonance
with $2878\pm 128$~MeV in $0^+ cs\bar{q}\bar{q}$ system \cite{tetraquark=Agaev}. There are also some work disfavored these assignments.
T.~J.~Burns {\em et~al.} interpreted the $X(2900)$ as a triangle cusp effect arising from $D^*\bar{K^*}$ and $D_{1}\bar{K}$ interactions \cite{molecular=baners}.
Based on an extended relativized quark model, the work \cite{tetraquark=Lu} found  four stable resonances, $2765$~MeV, $3055$~MeV, $3152$~MeV and $3396$~MeV,
and none of them could be the candidate of $X(2900)$ in the $0^+$ $cs\bar{q}\bar{q}$ system.

In fact, both molecular $D^*\bar{K^*}$ and diquark $cs\bar{q}\bar{q}$ configuration have energies near the mass of $X(2900)$.
Which structure is preferred should be determined by system dynamics. So the the structure mixing calculation is needed. Because of the high
energy of $X(2900)$, the combinations of excited states of $c\bar{q}$ and $s\bar{q}$ are possible. More important, these states will couple with the
decay channels, $D\bar{K}$, $D\bar{K}^*$ and $D^*\bar{K}$, do these states survive after the coupling? Due to the finite space used in the calculation,
a stability method has to be employed to identify the genuine resonance. In this paper, a structure mixing calculation of meson-meson and
diquark-antidiquark structures is performed in the framework of chiral quark model (ChQM), and the excited states of subclusters are included.
So four kinds of states with quantum numbers,
$00^{\pm}$ and $01^{\pm}$, are investigated. Due to lacking of orbit-spin force in our calculation, we use the symbol ``$^{2S+1}L_J$" denotes $P$-wave
excited states. So, $0^{-}$ and $1^{-}$ may be expressed as $^1P_1$, $^3P_J$ and $^5P_J$. To find the genuine resonance, real-scaling method \cite{RSM} is used.

The paper is organized as follows. In section II, the chiral quark model, real-scaling method and the wave-function of $cs\bar{q}\bar{q}$
systems are presented. The numerical results are given in Sec. III. The last section is devoted to the summary of the present work.

\section{Chiral quark model and wave function of $cs\bar{q}\bar{q}$ system} \label{wavefunction and chiral quark model}
The ChQM has been successful both in describing the hadron spectra and hadron-hadron interactions. The details of the model can be found in
Refs. \cite{quark-model-salamenca,quark-modle-our}. The Hamiltonian of CCQM consists of quarks mass, kinetic energy, and three kinds of potentials,
color confinement, one-gluon-exchange and Goldstone boson exchange. The Hamiltonian for four-quark system is written as,
\begin{eqnarray}
H &=& \sum_{i=1}^4
m_i+\frac{p_{12}^2}{2\mu_{12}}+\frac{p_{34}^2}{2\mu_{34}}
  +\frac{p_{1234}^2}{2\mu_{1234}}  \nonumber \\
  & & +\sum_{i<j=1}^4 \left( V_{ij}^{G}+V_{ij}^{C}+\sum_{\chi=\pi,K,\eta, \sigma} V_{ij}^{\chi} \right), \nonumber\\
\end{eqnarray}
Where $m$ is the constituent masse of quark (antiquark), and $\mu$ is the reduced masse of two interacting quarks or
quark-clusters.
\begin{eqnarray}
\mu_{ij}&=&\frac{m_{{i}}  m_{{j}}}{m_{{i}} + m_{{j}}},~ \mu_{1234}=\frac{(m_1+m_2)(m_3+m_4)}{m_1+m_2+m_3+m_4}, \\ \nonumber
p_{ij} &=&\frac{m_jp_i-m_ip_j}{m_i+m_j}, \nonumber \\
p_{1234}&=&\frac{(m_3+m_4)p_{12}-(m_1+m_2)p_{34}}{m_1+m_2+m_3+m_4}
\end{eqnarray}

The color confinement, the effective smeared one-gluon exchange interaction and the Goldstone boson exchange potential are,
\begin{equation}
V_{ij}^{C}= ( -a_{c} r_{ij}^{2}-\Delta) \boldsymbol{\lambda}_i^c \cdot \boldsymbol{\lambda}_j^c .
\end{equation}
\begin{eqnarray}
 V_{ij}^{G}&=& \frac{\alpha_s}{4} \boldsymbol{\lambda}_i^c \cdot \boldsymbol{\lambda}_{j}^c
\left[\frac{1}{r_{ij}}-\frac{2\pi}{3m_im_j}\boldsymbol{\sigma}_i\cdot
\boldsymbol{\sigma}_j
  \delta(\boldsymbol{r}_{ij})\right]   \\
 & &  \delta{(\boldsymbol{r}_{ij})}=\frac{e^{-r_{ij}/r_0(\mu_{ij})}}{4\pi r_{ij}r_0^2(\mu_{ij})},~~ r_{0}(\mu_{ij})=\frac{r_0}{\mu_{ij}}.
  \nonumber
\end{eqnarray}
\begin{eqnarray}
V_{ij}^{\pi}&=& \frac{g_{ch}^2}{4\pi}\frac{m_{\pi}^2}{12m_im_j}
  \frac{\Lambda_{\pi}^2}{\Lambda_{\pi}^2-m_{\pi}^2}m_\pi v_{ij}^{\pi}
  \sum_{a=1}^3 \lambda_i^a \lambda_j^a,  \nonumber \\
V_{ij}^{K}&=& \frac{g_{ch}^2}{4\pi}\frac{m_{K}^2}{12m_im_j}
  \frac{\Lambda_K^2}{\Lambda_K^2-m_{K}^2}m_K v_{ij}^{K}
  \sum_{a=4}^7 \lambda_i^a \lambda_j^a,  \nonumber \\
V_{ij}^{\eta} & = &
\frac{g_{ch}^2}{4\pi}\frac{m_{\eta}^2}{12m_im_j}
\frac{\Lambda_{\eta}^2}{\Lambda_{\eta}^2-m_{\eta}^2}m_{\eta}
v_{ij}^{\eta} \nonumber \\
 && \left[\lambda_i^8 \lambda_j^8 \cos\theta_P
 - \lambda_i^0 \lambda_j^0 \sin \theta_P \right],  \nonumber \\
V_{ij}^{\sigma}&=& -\frac{g_{ch}^2}{4\pi}
\frac{\Lambda_{\sigma}^2}{\Lambda_{\sigma}^2-m_{\sigma}^2}m_\sigma
\left[
 Y(m_\sigma r_{ij})-\frac{\Lambda_{\sigma}}{m_\sigma}Y(\Lambda_{\sigma} r_{ij})\right] \nonumber \\
 v_{ij}^{\chi} & = & \left[ Y(m_\chi r_{ij})-
\frac{\Lambda_{\chi}^3}{m_{\chi}^3}Y(\Lambda_{\chi} r_{ij})
\right]
\boldsymbol{\sigma}_i \cdot\boldsymbol{\sigma}_j, \nonumber \\
& & Y(x)  =   e^{-x}/x .
\end{eqnarray}
In the above formula, $\boldsymbol{\sigma}$ are the $SU(2)$ Pauli matrices; $\boldsymbol{\lambda}$, $\boldsymbol{\lambda}^{c}$
are $SU(3)$ flavor, color Gell-Mann matrices, respectively; $\alpha_{s}$ is an effective scale-dependent running coupling,
\begin{equation}
 \alpha_s(\mu_{ij})=\frac{\alpha_0}{\ln\left[(\mu_{ij}^2+\mu_0^2)/\Lambda_0^2\right]},
\end{equation}

For determining all the parameters, the first step of our study is to accommodate all the mesons, from light to heavy, taking into account only
a quark-antiquark component.
All of parameters are shown in Table~\ref{modelparameters} and the obtained masses of the mesons involved in the present calculation are
listed in Table \ref{mesonmass}.

\begin{table}[t]
\begin{center}
\caption{Quark Model Parameters ($m_{\pi}=0.7$ fm, $m_{\sigma}=3.42$ fm, $m_{\eta}=2.77$ fm, $m_{K}=2.51$ fm).\label{modelparameters}}
\begin{tabular}{cccc}
\hline\noalign{\smallskip}
Quark masses   &$m_u=m_d$(MeV)     &313  \\
               &$m_{s}$(MeV)         &536  \\
               &$m_{c}$(MeV)         &1728 \\
               &$m_{b}$(MeV)         &5112 \\
\hline
Goldstone bosons   &$\Lambda_{\pi}=\Lambda_{\sigma}(fm^{-1})$     &4.2  \\
                   &$\Lambda_{\eta}=\Lambda_{K}(fm^{-1})$     &5.2  \\
                   &$g_{ch}^2/(4\pi)$                &0.54  \\
                   &$\theta_p(^\circ)$                &-15 \\
\hline
Confinement        &$a_{c}$(MeV)     &101 \\
                   &$\Delta$(MeV)       &-78.3 \\
                   &$\mu_{c}$(MeV)       &0.7 \\
\hline
OGE                 & $\alpha_{0}$        &3.67 \\
                   &$\Lambda_{0}(fm^{-1})$ &0.033 \\
                  &$\mu_0$(MeV)    &36.976 \\
                   &$\hat{r}_0$(MeV)    &28.17 \\
\hline
\end{tabular}
\end{center}
\end{table}

\begin{table}[]
\caption{ \label{mesonmass} Meson spectrum (unit: MeV).}
\begin{tabular}{ccccc}
\hline\noalign{\smallskip}
        &~~ $D$    &~~ $D^*$  &~~  $D_J$  &~~ $D_1$\\ \hline
QM      &~~ 1862.6 &~~ 1980.5 &~~ 2454.7  &~~ 2448.1 \\
exp     &~~ 1867.7 &~~ 2008.9 &~~  2420.0 &~~ 2420.0 \\
\hline
        &~~ $K$    &~~ $K^*$  &~~  $K_J$  &~~ $K_1$\\ \hline
QM      &~~ 493.9  &~~ 913.6  &~~  1423.0 &~~ 1400.0 \\
exp     &~~ 495.0  &~~ 892.0  &~~  1430.0 &~~ 1427.0 \\
\hline
\end{tabular}
\end{table}

\subsection{The wave-function of $cs\bar{q}\bar{q}$ system}

Generally, $cs\bar{q}\bar{q}$ system has two interesting structures, meson-meson and diquark-antidiquark, and the wave function of each structure
all consists of four parts: orbit, spin, flavor and color wave functions. In addition, the wave function of each part is constructed
by coupling two sub-clusters wave functions. Thus, the wave function for each channel will be the tensor product of orbit
($|R_{i}\rangle$), spin ($|S_{j}\rangle$), color ($|C_{k}\rangle$) and flavor ($|F_{l}\rangle$) components,
\begin{equation}\label{bohanshu}
|ijkl\rangle={\cal A} |R_{i}\rangle\otimes|S_{j}\rangle\otimes|C_{k}\rangle\otimes|F_{l}\rangle
\end{equation}
${\cal A}$ is the antisymmetrization operator.

\subsubsection{Orbit wave function}
The total wave function consists of two sub-clusters orbit wave functions and the relative motion wave function between two sub-clusters.
\begin{eqnarray}\label{spatialwavefunctions}
|R_i\rangle & = & \left[ \left[ \Psi_{l_1}({\bf r}_{12})\Psi_{l_2}({\bf r}_{34})\right]_{l_{12}} \Psi_{l_3}({\bf r}_{1234})\right]_L.
\end{eqnarray}
The negative $P$ parity calls for angular momentum $l_i=1$, and we set only one $P$-wave angular momentum in a sub-cluster. In this assignment,
we may take combination of $l_1=1,l_2=0,l_3=0$ as "$|R_{1}\rangle$", and  $l_1=0,l_2=1,l_3=0$ as "$|R_{2}\rangle$". For the positive $P$ parity we
set all angular momentum to 0, $l_1=0,l_2=0,l_3=0$, which is denoted as "$|R_{3}\rangle$".

In GEM, the radial part of spatial wave function is expanded by Gaussians:
\begin{subequations}
\label{radialpart}
\begin{align}
\Psi_{lm}(\mathbf{r}) & = \sum_{n=1}^{n_{\rm max}} c_{n}\psi^G_{nlm}(\mathbf{r}),\\
\psi^G_{nlm}(\mathbf{r}) & = N_{nl}r^{l}
e^{-\nu_{n}r^2}Y_{lm}(\hat{\mathbf{r}}),
\end{align}
\end{subequations}
where $N_{nl}$ are normalization constants,
\begin{align}
N_{nl}=\left[\frac{2^{l+2}(2\nu_{n})^{l+\frac{3}{2}}}{\sqrt{\pi}(2l+1)}
\right]^\frac{1}{2}.
\end{align}
$c_n$ are the variational parameters, which are determined dynamically. The Gaussian size parameters are chosen
according to the following geometric progression
\begin{equation}\label{gaussiansize}
\nu_{n}=\frac{1}{r^2_n}, \quad r_n=r_1a^{n-1}, \quad
a=\left(\frac{r_{n_{\rm max}}}{r_1}\right)^{\frac{1}{n_{\rm
max}-1}}.
\end{equation}
GEM has been successfully used in the calculation for $^4H_e$ and other few-body systems, its precision is similar with other methods,
such as Faddeev method, SMV method, HH method and so on \cite{GEM}. The advantage of GEM is that it converges rather fast.

\subsubsection{Spin wave function}
Because of no difference between spin of quark and antiquark, the meson-meson structure has the same total spin as
the diquark-antidiquark structure. 
\begin{eqnarray}
|S_{1}\rangle & = & \chi_{0}^{\sigma1}=\chi_{00}^{\sigma}\chi_{00}^{\sigma}, \nonumber \\
|S_{2}\rangle & = & \chi_{0}^{\sigma2}= \sqrt{\frac{1}{3}}(\chi_{11}^{\sigma}\chi_{1-1}^{\sigma}
  -\chi_{10}^{\sigma}\chi_{10}^{\sigma}+\chi_{1-1}^{\sigma}\chi_{11}^{\sigma}), \nonumber \\
|S_{3}\rangle & = & \chi_{1}^{\sigma1}=\chi_{00}^{\sigma}\chi_{11}^{\sigma}, \\
|S_{4}\rangle & = & \chi_{1}^{\sigma2}=\chi_{11}^{\sigma}\chi_{00}^{\sigma}, \nonumber \\
|S_{5}\rangle & = & \chi_{1}^{\sigma3}=\frac{1}{\sqrt{2}}(\chi_{11}^{\sigma}\chi_{10}^{\sigma}-\chi_{10}^{\sigma}\chi_{11}^{\sigma}),\nonumber \\
|S_{6}\rangle & = & \chi_{2}^{\sigma1}=\chi_{11}^{\sigma}\chi_{11}^{\sigma}. \nonumber
\end{eqnarray}
Where the subscript of "$\chi_{S}^{\sigma j}$" denotes total spin of the tretraquark, and the superscript is the index of
the spin function with fixed $S$.

\subsubsection{Flavor wave function}
 The total flavor wave functions can be written as,
\begin{eqnarray}
|F_{1}\rangle & = & \frac{1}{\sqrt{2}}\left( c\bar{u}s\bar{d}- c\bar{d}s\bar{u} \right), \nonumber\\
|F_{2}\rangle & = & \frac{1}{\sqrt{2}}\left( cs\bar{u}\bar{d}- cs\bar{d}\bar{u} \right)
\end{eqnarray}
Where $|F_{1}\rangle$ means molecular flavor while $|F_{2}\rangle$ means diaquark flavor.

\subsubsection{Color wave function}
The colorless tetraquark system has four color structures, including $1\otimes1$, $8\otimes8$, $3\otimes \bar{3}$ and $6\otimes \bar{6}$,
\begin{eqnarray}
|C_{1}\rangle & = & \chi_{1\otimes1}^{m1}=\frac{1}{\sqrt{9}}(\bar{r}r\bar{r}r+\bar{r}r\bar{g}g+\bar{r}r\bar{b}b
   +\bar{g}g\bar{r}r+\bar{g}g\bar{g}g \nonumber \\
  & + & \bar{g}g\bar{b}b+\bar{b}b\bar{r}r+\bar{b}b\bar{g}g+\bar{b}b\bar{b}b), \nonumber \\
|C_{2}\rangle & = & \chi_{8\otimes8}^{m2}=\frac{\sqrt{2}}{12}(3\bar{b}r\bar{r}b+3\bar{g}r\bar{r}g+3\bar{b}g\bar{g}b
   +3\bar{g}b\bar{b}g \nonumber \\
 &+ & 3\bar{r}g\bar{g}r+ 3\bar{r}b\bar{b}r+2\bar{r}r\bar{r}r+2\bar{g}g\bar{g}g+2\bar{b}b\bar{b}b-\bar{r}r\bar{g}g \nonumber \\
&-& \bar{g}g\bar{r}r-\bar{b}b\bar{g}g-\bar{b}b\bar{r}r-\bar{g}g\bar{b}b-\bar{r}r\bar{b}b). \\
|C_{3}\rangle & = & \chi^{d1}_{\bar{3}\otimes 3} =\frac{\sqrt{3}}{6}(rg\bar{r}\bar{g}-rg\bar{g}\bar{r}+gr\bar{g}\bar{r}
    -gr\bar{r}\bar{g}+rb\bar{r}\bar{b}, \nonumber \\
&- & rb\bar{b}\bar{r}+br\bar{b}\bar{r}-br\bar{r}\bar{b}+gb\bar{g}\bar{b}-gb\bar{b}\bar{g}+bg\bar{b}\bar{g}-bg\bar{g}\bar{b}), \nonumber \\
|C_{4}\rangle & = & \chi^{d2}_{6\otimes \bar{6}}=\frac{\sqrt{6}}{12}(2rr\bar{r}\bar{r}+2gg\bar{g}\bar{g}+2bb\bar{b}\bar{b}
    +rg\bar{r}\bar{g} \nonumber \\
&+ &rg\bar{g}\bar{r}+gr\bar{g}\bar{r}+gr\bar{r}\bar{g}+rb\bar{r}\bar{b}+rb\bar{b}\bar{r}+br\bar{b}\bar{r} \nonumber \\
&+ &br\bar{r}\bar{b}+gb\bar{g}\bar{b}+gb\bar{b}\bar{g}+bg\bar{b}\bar{g}+bg\bar{g}\bar{b}).\nonumber
\end{eqnarray}
To write down the wave functions easily for each structure, the different orders of the particles are used. However, when coupling the
different structure, the same order of the particles should be used.

\subsubsection{Total wave function}
In the present work, we investigated all possible candidates of $X(2900)$ in the $cs\bar{q}\bar{q}$ system.
The antisymmetrization operators are different for different structures.
For $cs\bar{q}\bar{q}$ system, the antisymmetrization operator becomes
\begin{equation}
{\cal A}=1-(24)
\end{equation}
for meson-meson structure, and
\begin{equation}
{\cal A}=1-(34)
\end{equation}
for diquark-antidiquark structure. After applying the antisymmetrization operator, some wave function will vanish, which means
that the states are forbidden. All of allowed channels are listed in Table \ref{channel}. The subscript ``8" denotes color octet subcluster,
the superscript of diquark/antidiquark is the spin of the subcluster, and the subscript is the color representation of subcluster, $3$,
$\bar{3}$, $6$ and $\bar{6}$ denote color triplet, anti-triplet, sextet and anti-sextet.
\begin{table}[ht]
\centering
\fontsize{10}{10}\selectfont
\caption{All of allowed channels(we use $|ijkl>$ donates different states. The "i" means different angular momentum combinations; The "j" means different spin channels; The "k" means different flavor channels; The "l" means different color configurations)}.\label{channel}
\begin{tabular}{cccccc}
\hline \hline
 \multicolumn{6}{c}{$cs\bar{q}\bar{q}$}  \\ \hline
 $ |ijkl\rangle $ & $^3P_J                                 $&$ |ijkl\rangle $ & $^1P_1 $&                                  $ |ijkl\rangle $& $^5P_J$ \\ \hline
 $ |1311\rangle $ & $ D_1 \bar{K^*}                        $&$ |1111\rangle $ & $ D_1 \bar{K}                          $&$ |1611\rangle $ & $ D_J \bar{K^*}                        $\\
 $ |1312\rangle $ & $[D_1]_8[\bar{K^*}]_8                  $&$ |1112\rangle $ & $[D_1]_8[\bar{K}]_8                    $&$ |1612\rangle $ & $[D_J]_8[\bar{K^*}]_8                  $\\
 $ |1411\rangle $ & $ D_J \bar{K}                          $&$ |1211\rangle $ & $ D_J \bar{K^*}                        $&$ |2611\rangle $ & $ D^* \bar{K_J}                        $\\
 $ |1412\rangle $ & $[D_J]_8[\bar{K}]_8                    $&$ |1212\rangle $ & $[D_J]_8[\bar{K^*}]_8                  $&$ |2612\rangle $ & $[D^*]_8[\bar{K_J}]_8                  $\\
 $ |1511\rangle $ & $ D_J \bar{K^*}                        $&$ |2111\rangle $ & $ D \bar{K_1}                          $&$ |1622\rangle $ & $ [cs]_6^1 [\bar{q}\bar{q}]_{\bar{6}}^1$\\
 $ |1512\rangle $ & $[D_J]_8[\bar{K^*}]_8                  $&$ |2112\rangle $ & $[D]_8[\bar{K_1}]_8                    $&$ |2621\rangle $ & $ [cs]_3^1 [\bar{q}\bar{q}]_{\bar{3}}^1$\\ \cline{5-6}
 $ |2311\rangle $ & $ D \bar{K_J}                          $&$ |2211\rangle $ & $ D^* \bar{K_J}                        $&$ |ijkl\rangle $ & $1^{+}$                                 \\ \cline{5-6}
 $ |2312\rangle $ & $[D]_8[\bar{K_J}]_8                    $&$ |2212\rangle $ & $[D^*]_8[\bar{K_J}]_8                  $&$ |0311\rangle $ & $ D \bar{K^*}                            $\\
 $ |2411\rangle $ & $ D^* \bar{K_1}                        $&$ |1123\rangle $ & $ [cs]_3^0 [\bar{q}\bar{q}]_{\bar{3}}^0$&$ |0312\rangle $ & $[D]_8[\bar{K^*}]_8                      $\\
 $ |2412\rangle $ & $[D^*]_8[\bar{K_1}]_8                  $&$ |1124\rangle $ & $ [cs]_6^1 [\bar{q}\bar{q}]_{\bar{6}}^1$&$ |0411\rangle $ & $ D^* \bar{K}                            $\\
 $ |2511\rangle $ & $ D^* \bar{K_J}                        $&$ |2124\rangle $ & $ [cs]_6^0 [\bar{q}\bar{q}]_{\bar{6}}^0$&$ |0412\rangle $ & $[D^*]_8[\bar{K}]_8                      $\\
 $ |2512\rangle $ & $[D^*]_8[\bar{K_J}]_8                  $&$ |2223\rangle $ & $ [cs]_3^1 [\bar{q}\bar{q}]_{\bar{3}}^1$&$ |0511\rangle $ & $ D^* \bar{K^*}                            $\\ \cline{3-4}
 $ |1324\rangle $ & $ [cs]_6^0 [\bar{q}\bar{q}]_{\bar{6}}^1$&$ |ijkl\rangle $ & $0^{+}                                 $&$ |0512\rangle $ & $[D^*]_8[\bar{K^*}]_8                      $\\ \cline{3-4}
 $ |1423\rangle $ & $ [cs]_3^1 [\bar{q}\bar{q}]_{\bar{3}}^0$&$ |0111\rangle $ & $ D \bar{K}                            $&$ |0324\rangle $ & $ [cs]_6^0 [\bar{q}\bar{q}]_{\bar{6}}^1$\\
 $ |1524\rangle $ & $ [cs]_6^1 [\bar{q}\bar{q}]_{\bar{6}}^1$&$ |0112\rangle $ & $[D]_8[\bar{K}]_8                      $&$ |0423\rangle $ & $ [cs]_3^1 [\bar{q}\bar{q}]_{\bar{3}}^0$\\
 $ |2323\rangle $ & $ [cs]_3^0 [\bar{q}\bar{q}]_{\bar{3}}^1$&$ |0211\rangle $ & $ D^* \bar{K^*}                        $&$ |0524\rangle $ & $ [cs]_6^1 [\bar{q}\bar{q}]_{\bar{6}}^1$\\
 $ |2424\rangle $ & $ [cs]_6^1 [\bar{q}\bar{q}]_{\bar{6}}^0$&$ |0212\rangle $ & $[D^*]_8[\bar{K^*}]_8                  $\\
 $ |2523\rangle $ & $ [cs]_3^1 [\bar{q}\bar{q}]_{\bar{3}}^1$&$ |0123\rangle $ & $ [cs]_3^0 [\bar{q}\bar{q}]_{\bar{3}}^0$\\
                                                           &&$ |0223\rangle $ & $ [cs]_6^1 [\bar{q}\bar{q}]_{\bar{6}}^1$\\
  \hline\hline
\end{tabular}
\end{table}

\section{Result}
In this section, we present our numerical results. In the calculation of $cs\bar{q}\bar{q}$ system, two structures, meson-meson structure and
diquark-antidiquark structure and their coupling are considered. Due to the mass of $X(2900)$ is larger than the threshold of $cs\bar{q}\bar{q}$ system,
the possible candidates must be resonance states. To check that whether the candidates survive the coupling to the open channels, $D\bar{K}$, $D\bar{K}^*$
and $D^*\bar{K}$, The real-scaling method (RSM) was employed to test stability of these candidates.

\subsection{Possible candidates of $X(2900)$}

In $J^P=0^{+}$ $cs\bar{q}\bar{q}$ system, there are four channels in meson-meson structure and two channels in diquark-antidiquark structure
(see Table \ref{channel}). The lowest eigen-energy of each channel is given in the second column of Table \ref{0+E}. The eigen-energies of
full channel coupling are shown in the rows which are marked ``c.c", and the percentages in the table stand for percent of each channel in the
eigen-states with corresponding energies. The two lowest eigen-energies and the eigen-energies near 2900 MeV are given.
In the channel coupling calculation we get four energy levels, $E_1(2836)$, $E_2(2896)$, $E_3(2906)$ and $E_4(2936)$, which could be the candidates
of $X(2900)$. However, the eigen-state with $E_2(2896)$ has almost 89\% of $D^*\bar{K}^*$, and the energy is higher than its threshold, 2894 MeV,
and the single-channel calculation of $D^*\bar{K}^*$ reveals the state is unbound, so it should be a $D^*\bar{K^*}$ scattering state rather than
a new resonance. Thus, $X_0(2900)$ may not be $0^{+} D^*\bar{K}^*$ in our calculation.
The other candidates all have dominant meson-meson scattering states, for example the state with $E_3(2906)$ has $58.8\%$ $D^*\bar{K}^*$ scattering state
and $24\%$ $D\bar{K}$ scattering state while diquark structure has $15\%$. One has to check the stability of these states if assign these candidates
to $X_0(2900)$.

Due to three combinations of spin in the $J^P=1^{+}$ $cs\bar{q}\bar{q}$ system, more channels are listed in the table\ref{0+E}. Consequently,
five energy levels near $X(2900)$, $E_5(2857)$, $E_6(2896)$, $E_7(2904)$, $E_8(2920)$ and $E_9(2941)$ are founded in the channel coupling calculation.
Similar to the $J^P=0^{+}$ case, all these states are dominated by the meson-meson scattering states. On the other hand, the lowest energy of diquark
structure is $ [cs]_3^1 [\bar{q}\bar{q}]_{\bar{3}}^0 $ with $2690$~MeV which is not suitable to be candidate of $X(2900)$.
\begin{table}[ht]
\centering
\fontsize{7}{7}\selectfont
\caption{All of allowed channels(we use $|ijkl\rangle$ donates different states. The "i" means different angular momentum combinations; The "j" means different spin channels; The "k" means different flavor channels; The "l" means different color configurations)}.\label{0+E}
\begin{tabular}{ccccccccc}
\hline \hline
 \multicolumn{9}{c}{$0^{+}~cs\bar{q}\bar{q}$}  \\ \hline
                                        & s.c.   &1st     & 2nd    &...& 7th    & 8th    & 9st    & 10th  \\
$ D \bar{K}                            $&$2357.0$&$90.1\%$&$99.4\%$&...&$61.9\%$&$ 6.1\%$&$24.0\%$&$26.2\%$\\
$[D]_8[\bar{K}]_8                      $&$3098.2$&$ 0.3\%$&$ 0.0\%$&...&$ 1.5\%$&$ 0.4\%$&$ 0.9\%$&$ 4.2\%$\\
$ D^* \bar{K^*}                        $&$2895.8$&$ 0.5\%$&$ 0.0\%$&...&$ 0.3\%$&$88.8\%$&$58.8\%$&$28.9\%$\\
$[D^*]_8[\bar{K^*}]_8                  $&$2863.7$&$ 1.5\%$&$ 0.1\%$&...&$ 2.6\%$&$ 0.1\%$&$ 1.5\%$&$ 0.4\%$\\
$ [cs]_3^0 [\bar{q}\bar{q}]_{\bar{3}}^0$&$2656.5$&$ 6.9\%$&$ 0.1\%$&...&$ 7.3\%$&$ 0.3\%$&$10.1\%$&$ 0.9\%$\\
$ [cs]_6^1 [\bar{q}\bar{q}]_{\bar{6}}^1$&$2965.7$&$ 0.7\%$&$ 0.4\%$&...&$26.3\%$&$ 4.4\%$&$ 4.9\%$&$29.8\%$\\
   c.c.                                 &        &$2340.1$&$2358.9$&...&$2836.3$&$2896.7$&$2906.9$&$2935.8$\\
  \hline
  \multicolumn{9}{c}{$1^{+}~cs\bar{q}\bar{q}$}  \\ \hline
                                         & s.c.   & 1st    &...& 10th   & 11th   & 12th   & 13th   &14th    \\
$ D \bar{K^*}                           $&$2777.6$&$ 0.3\%$&...&$25.4\%$&$ 0.5\%$&$ 0.4\%$&$ 0.2\%$&$ 9.3\%$\\
$[D]_8[\bar{K^*}]_8                     $&$3111.8$&$ 0.4\%$&...&$ 3.4\%$&$ 1.1\%$&$ 1.2\%$&$ 1.7\%$&$ 0.4\%$\\
$ D^*\bar{K}                            $&$2475.3$&$87.4\%$&...&$34.2\%$&$30.1\%$&$55.2\%$&$63.7\%$&$54.8\%$\\
$[D^*]_8[\bar{K}]_8                     $&$3110.7$&$ 0.4\%$&...&$ 1.6\%$&$ 0.6\%$&$ 1.8\%$&$ 1.2\%$&$ 0.9\%$\\
$ D^* \bar{K^*}                         $&$2895.9$&$ 0.3\%$&...&$ 0.6\%$&$52.5\%$&$11.1\%$&$ 3.7\%$&$ 0.9\%$\\
$[D^*]_8[\bar{K^*}]_8                   $&$3005.0$&$ 1.1\%$&...&$ 0.9\%$&$ 1.5\%$&$ 2.8\%$&$ 3.4\%$&$ 4.6\%$\\
$ [cs]_6^0 [\bar{q}\bar{q}]_{\bar{6}}^1 $&$3112.3$&$ 0.1\%$&...&$ 0.4\%$&$ 1.2\%$&$ 7.3\%$&$ 4.8\%$&$19.7\%$\\
$ [cs]_3^1 [\bar{q}\bar{q}]_{\bar{3}}^0 $&$2690.6$&$ 9.8\%$&...&$ 0.5\%$&$ 1.2\%$&$ 0.8\%$&$ 4.5\%$&$ 5.5\%$\\
$ [cs]_6^1 [\bar{q}\bar{q}]_{\bar{6}}^1 $&$3040.4$&$ 0.1\%$&...&$16.1\%$&$11.4\%$&$19.4\%$&$17.0\%$&$ 3.9\%$\\
    c.c.                                 &        &$2464.3$&...&$2857.1$&$2896.3$&$2904.2$&$2920.3$&$2941.7$\\
  \hline
\end{tabular}
\end{table}
\begin{table}[tp]
\centering
\caption{Index of physical channels for $c\bar{s}s\bar{c}$ system.}\label{channel4}
\begin{tabular}{cccc}
\hline \hline
 \multicolumn{2}{c}{$^3P_J$}& \multicolumn{2}{c}{$^1P_1$} \\ \hline
$ D_1 \bar{K^*}                        ~~~$&$3363.6~~~$&$ D_1 \bar{K}                          ~~~$&$2943.3$\\
$[D_1]_8[\bar{K^*}]_8                  ~~~$&$3551.3~~~$&$[D_1]_8[\bar{K}]_8                    ~~~$&$3554.8$\\
$ D_J \bar{K}                          ~~~$&$2950.0~~~$&$ D_J \bar{K^*}                        ~~~$&$3369.5$\\
$[D_J]_8[\bar{K}]_8                    ~~~$&$3556.2~~~$&$[D_J]_8[\bar{K^*}]_8                  ~~~$&$3340.5$\\
$ D_J \bar{K^*}                        ~~~$&$3370.2~~~$&$ D \bar{K_1}                          ~~~$&$3264.4$\\
$[D_J]_8[\bar{K^*}]_8                  ~~~$&$3448.5~~~$&$[D]_8[\bar{K_1}]_8                    ~~~$&$3544.5$\\
$ D \bar{K_J}                          ~~~$&$3287.4~~~$&$ D^* \bar{K_J}                        ~~~$&$3404.8$\\
$[D]_8[\bar{K_J}]_8                    ~~~$&$3543.9~~~$&$[D^*]_8[\bar{K_J}]_8                  ~~~$&$3334.2$\\
$ D^* \bar{K_1}                        ~~~$&$3382.5~~~$&$ [cs]_3^0 [\bar{q}\bar{q}]_{\bar{3}}^0~~~$&$3036.1$\\
$[D^*]_8[\bar{K_1}]_8                  ~~~$&$3539.9~~~$&$ [cs]_6^1 [\bar{q}\bar{q}]_{\bar{6}}^1~~~$&$3279.5$\\
$ D^* \bar{K_J}                        ~~~$&$3405.4~~~$&$ [cs]_6^0 [\bar{q}\bar{q}]_{\bar{6}}^0~~~$&$3483.4$\\
$[D^*]_8[\bar{K_J}]_8                  ~~~$&$3434.9~~~$&$ [cs]_3^1 [\bar{q}\bar{q}]_{\bar{3}}^1~~~$&$3621.6$\\
$ [cs]_6^0 [\bar{q}\bar{q}]_{\bar{6}}^1~~~$&$3372.1~~~$&c.c.                                       &$2908.1$\\ \cline{3-4}
$ [cs]_3^1 [\bar{q}\bar{q}]_{\bar{3}}^0~~~$&$3037.3~~~$&    \multicolumn{2}{c}{$^5P_J$}                     \\ \cline{3-4}
$ [cs]_6^1 [\bar{q}\bar{q}]_{\bar{6}}^1~~~$&$3327.4~~~$& $ D_J \bar{K^*}                        $&$3370.3$\\
$ [cs]_3^0 [\bar{q}\bar{q}]_{\bar{3}}^1~~~$&$3625.5~~~$& $[D_J]_8[\bar{K^*}]_8                  $&$3653.3$\\
$ [cs]_6^1 [\bar{q}\bar{q}]_{\bar{6}}^0~~~$&$3477.1~~~$& $ D^* \bar{K_J}                        $&$3405.4$\\
$ [cs]_3^1 [\bar{q}\bar{q}]_{\bar{3}}^1~~~$&$3640.1~~~$& $[D^*]_8[\bar{K_J}]_8                  $&$3649.8$\\
c.c.                                       &$2941.0~~~$& $ [cs]_6^1 [\bar{q}\bar{q}]_{\bar{6}}^1$&$3416.2$\\
                                           &           & $ [cs]_3^1 [\bar{q}\bar{q}]_{\bar{3}}^1$&$3675.6$\\
                                           &           & c.c.            &              $3221.6$  \\
\hline
\end{tabular}
\end{table}

For the $P$-wave excited $cs\bar{q}\bar{q}$ system, the states are denoted as $^1P_1$, $^3P_J$ ($J=0,1,2$) and $^5P_J$ ($J=1,2,3$).
Because the present calculation only includes center force interaction, the states with the same spin are degenerate. For the reason the
$P$-wave $cs\bar{q}\bar{q}$ threshold $D_J \bar{K}$ is close to $X(2900)$, $X(2900)$ may be a molecular $D_J$-$\bar{K}$ state.
The structures coupling result are shown in Table \ref{channel4}, and we get two bounding states which can be the $X(2900)$ in Table \ref{channel4}.
The first state is $E_{10}(2908)$ in the $^1P_1$ $cs\bar{q}\bar{q}$, and it's mainly $D_1 \bar{K}$ with 35 MeV binding energy.
Because the energy of $D_1 \bar{K}$ is close to $ [cs]_3^0 [\bar{q}\bar{q}]_{\bar{3}}^0$, the coupling effect will plays important role in the form of
$E_{10}(2908)$. On the other hand, there is an other shallow bound state, $E_{11}(2941)$, existing in the $^3P_J$ $cs\bar{q}\bar{q}$ with several
MeVs binding energies. It should be molecular $D_J\bar{K}$.

\subsection{Candidates of $X(2900)$}
In last subsection, we get 11 possible candidates of $X(2900)$ in two structures, di-meson and diquark-antidiquark and their coupling.
However, The LHCb only found two resonances near the $2900$~MeV, and the number of candidates may be too rich for $X(2900)$. There are two reason
why our model provides so many candidates. Firstly, we take two different structures into our calculation at the same time, which result in molecular
energies and diquark energies filling our coupling energies. Secondly, it's impossible for theoretical work to expand their calculation space to
bigger infinitely, and the limited calculation space always offers false resonances. As consequences, to see if these states are genuine resonances
or not, the real-scaling method \cite{RSm} is employed. In this method, the Gaussian size parameters $r_n$ for the basis functions between two
sub-clusters for the color-singlet channels are scaled by multiplying a factor $\alpha$, i.e. $r_n \rightarrow \alpha r_n$. Then, any continuum
state will fall off towards its threshold, while a compact resonant state should not be affected by the variation of $\alpha$.

\begin{figure}[htp]
\begin{center}
\centerline{\epsfxsize=9cm\epsffile{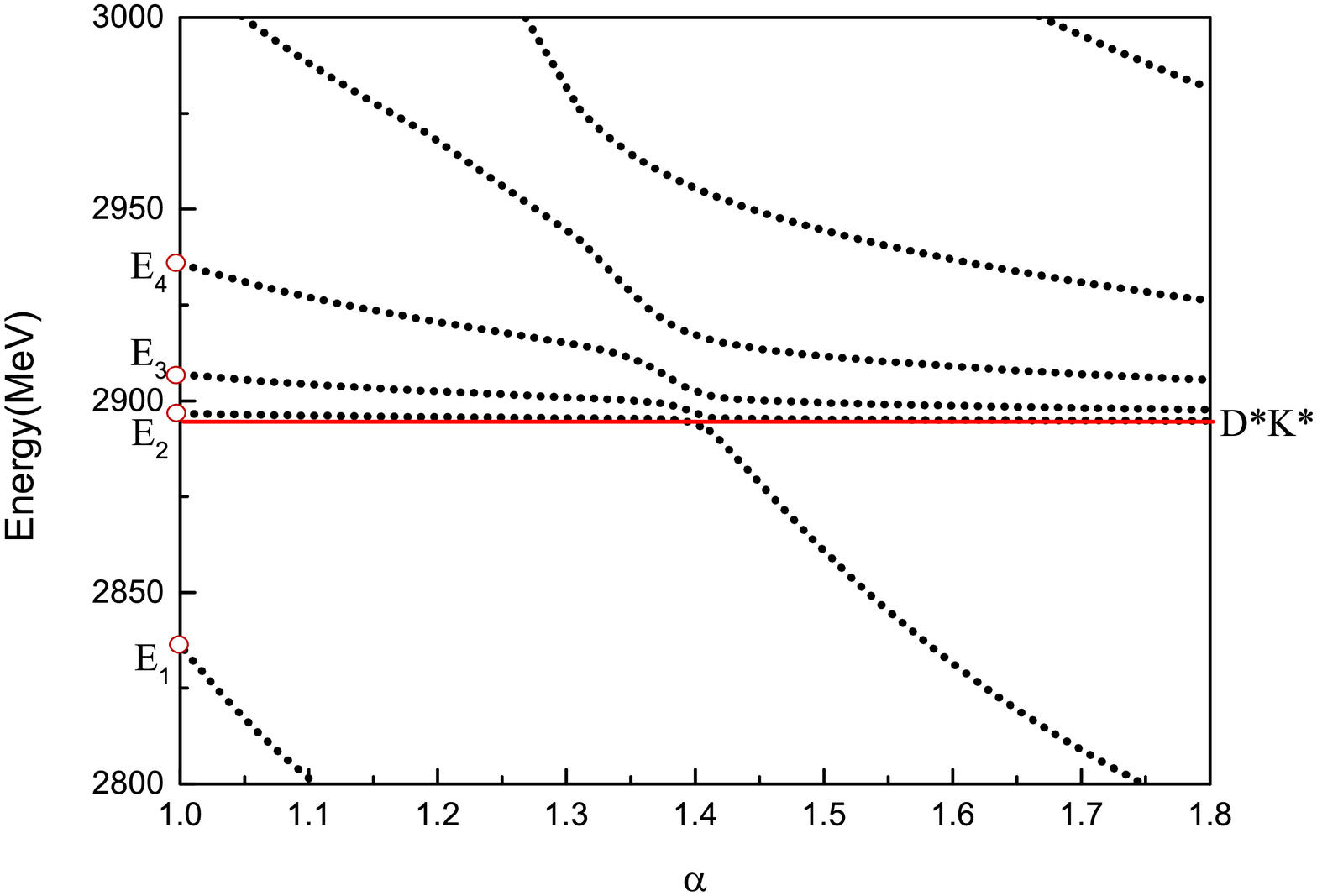}}
\caption{ Energy spectrum of $00^{+}$ states. }\label{00}
\end{center}
\end{figure}

\begin{figure}[htp]
\begin{center}
\centerline{\epsfxsize=9cm\epsffile{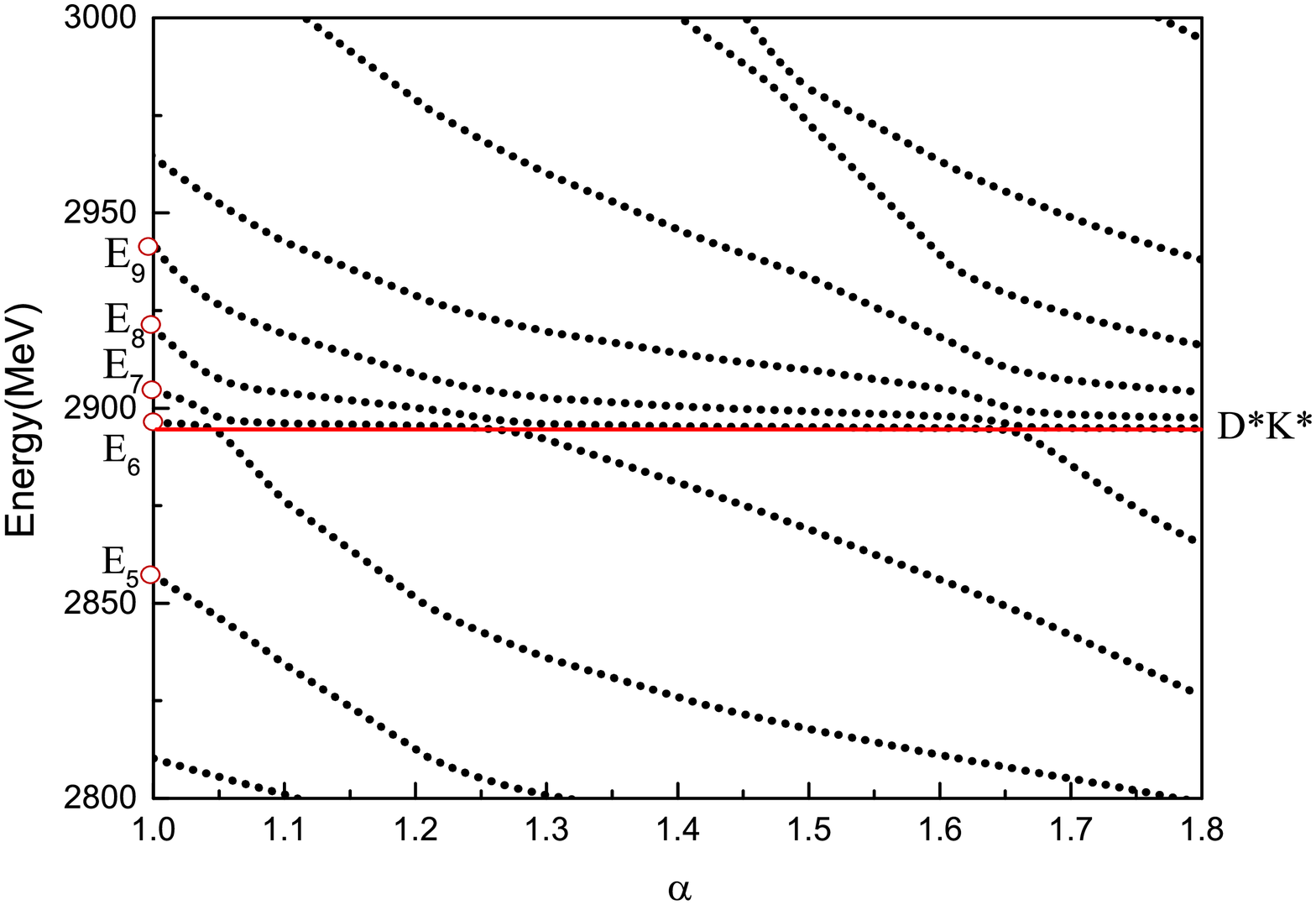}}
\caption{ Energy spectrum of $01^{+}$ states. }\label{01}
\end{center}
\end{figure}

\begin{figure}[htp]
\begin{center}
\centerline{\epsfxsize=9cm\epsffile{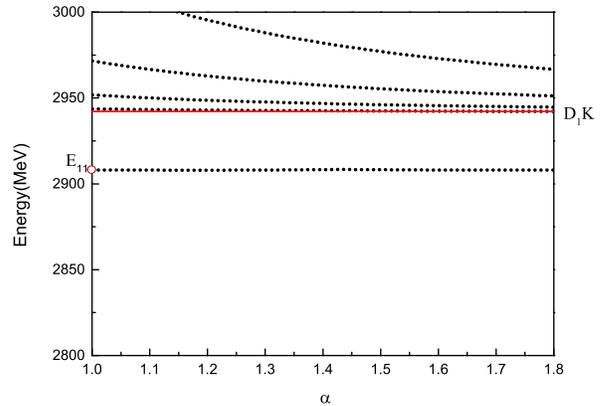}}
\caption{ Energy spectrum of $^1P_1$ states. }\label{1p1}
\end{center}
\end{figure}

\begin{figure}[htp]
\begin{center}
\centerline{\epsfxsize=9cm\epsffile{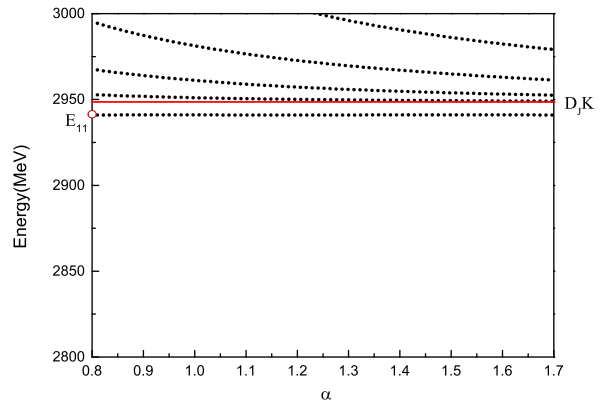}}
\caption{ Energy spectrum of $^3P_J$ states. }\label{3p1}
\end{center}
\end{figure}

The results for positive parity are shown in Figs.~\ref{00} and \ref{01} while negative parity are shown in Figs.~\ref{1p1} and \ref{3p1}.
Because we only focus on states with energy from 2800~MeV to 3000~MeV, other states are omitted in the figures.
We can see that only $D^*\bar{K^*}$ state appears in the Figs.~\ref{00} and \ref{01} which is marked with a red line. In the Figs.~\ref{00},
the resonance $E_1(2836)$ rapidly falls to lowest threshold, $D\bar{K}$ with the spaces getting bigger,
and both $E_3(2906)$ and $E_4(2936)$ would fall off towards $D^*\bar{K^*}$ threshold. However, there is a avoid-crossing structure around $\alpha=1.4$, but the pattern is not repeated, large $\alpha$ region is needed to make sure the
appearance of the resonance structure. Similarly, the $01^{+}$ resonances, $E_7(2904)$, $E_8(2920)$ and $E_9(2941)$, also have the same behave, and
they are also un-observable resonances. As for the resonance, $E_5(2857)$ should be an other scattering state and would decay to $D\bar{K^*}$ and
$D\bar{K}$ threshold. These falling curves may indicate that the assignment of positive parity may not suitable for the $X(2900)$.

Let's turn to negative parity shown in Figs.~\ref{1p1} and \ref{3p1}. Due to the $P$-wave excited threshold close to $X(2900)$, we only focus
on states with binding energies, $E_{10}(2908)$ and $E_{11}(2941)$. And the two bound states are very stable against to decay to threshold $D_J\bar{K}$.
They can decay to $D\bar{K}$ with $P$-wave between two mesons, it is expected that this decay is not too large.
In this way, we may assign $X_1(2900)$ as $01^{-}(^1P_1)$ $cs\bar{q}\bar{q}$. For the reason without orbit-spin interaction we can not determine
$J$ quantum of $^3P_J$ $cs\bar{q}\bar{q}$. Because of $X_0(2900)$ is strongly couple to $D\bar{K}$, it is not suitable
to assign $X_0(2900)$ as $^3P_J$ $cs\bar{q}\bar{q}$.

\section{Summary}

In the framework of the chiral constituent quark model, we study systematically $cs\bar{q}\bar{q}$ states to find the candidates of $X(2900)$.
Not only molecular structure but also diquark-antidiquark, with all possible color, flavor, spin configurations are taken into account. By means of
structures coupling, we get many states with energies around 2900 MeV in the positive parity $cs\bar{q}\bar{q}$ system. Real-scaling method,
a powerful method for identifying the genuine resonance, is used to test the stabilities of these states, and the results show that no resonances
could be observable below the threshold of $D^*\bar{K}^*$, On the other hand, the pictures of $P$-wave
excited $cs\bar{q}\bar{q}$ system show two stable bound states, $D_1\bar{K}$ and $D_J\bar{K}$, which could be the candidate of $X(2900)$.
In the absence of spin-orbit interaction, we only give the result of $^1P_1$ and $^3P_J$. we assign the $^1P_1$ $cs\bar{q}\bar{q}$ as $X_1(2900)$ with $IJ^{P}=01^{-}$. Because of $X_0(2900)$ is strongly couple to $D\bar{K}$, it is not suitable to assign $X_0(2900)$ as $^3P_J$ $cs\bar{q}\bar{q}$. When the spin-orbit and tensor interactions are included,
the $^1P_1$ and $^3P_J$ will be mixed up and the eigen values would vary slightly and would not change our results.


\begin{thebibliography}{99}

  \bibitem{Belle-3872}
  S.K.~Choi et al, (Belle Collaboration),
  Phys.\ Rev.\ Lett {\bf 91}, 262001 (2003).

  \bibitem{CDF-3872}
  D.~Acosta et al, (CDFII Collaboration),
  Phys.\ Rev.\ Lett {\bf 93}, 072001 (2004).

  \bibitem{D0-3872}
  V.M.~Abazov et al, (D0 Collaboration),
  Phys.\ Rev.\ Lett {\bf 93}, 162002 (2004).

  \bibitem{Babar-3872}
  B.~Aubert et al, (BaBar Collaboration),
  Phys.\ Rev.\ D {\bf 71}, 071103 (2005).

\bibitem{3872-tan}
  Y.~Tan and J.~Ping,
  Phys.\ Rev.\ D {\bf 100}, no. 3, 034022 (2019)
  doi:10.1103/PhysRevD.100.034022
  [arXiv:1906.09690 [hep-ph]].

 \bibitem{D0-5568}
  V.~M.~Abazov {\it et al.} [D0 Collaboration],
  Phys.\ Rev.\ Lett.\  {\bf 117}, 022003 (2016).


 \bibitem{Lhcb-5568}
  The LHCb Collaboration [LHCb Collaboration],
  LHCb-CONF-2016-004, CERN-LHCb-CONF-2016-004.

\bibitem{PLB760} T. J. Burns, E. S. Swanson, Phys. Lett. B {\bf 760}, 627 (2016).

\bibitem{5568-1}
  X.~Chen and J.~Ping,
  Eur.\ Phys.\ J.\ C {\bf 76}, no. 6, 351 (2016)
  doi:10.1140/epjc/s10052-016-4210-x
  [arXiv:1604.05651 [hep-ph]].


 \bibitem{5568-huang}
  H.~Huang and J.~Ping,
  Eur.\ Phys.\ J.\ C {\bf 79}, 556 (2019)


\bibitem{Lhcb-2900}
  R.~Aaij {\it et al.} [LHCb Collaboration],
  arXiv:2009.00025 [hep-ex].

\bibitem{Lhcb-2900-2}
  R.~Aaij {\it et al.} [LHCb Collaboration],
  arXiv:2009.00026 [hep-ex].

\bibitem{molecular=baners}
  T.~J.~Burns and E.~S.~Swanson,
  arXiv:2008.12838 [hep-ph].

\bibitem{molecular=genglisheng}
  Y.~Huang, J.~X.~Lu, J.~J.~Xie and L.~S.~Geng,
  arXiv:2008.07959 [hep-ph].

\bibitem{molecular=genglisheng2}
  M.~Z.~Liu, J.~J.~Xie and L.~S.~Geng,
  arXiv:2008.07389 [hep-ph].

\bibitem{molecular=Hejun}
  J.~He and D.~Y.~Chen,
  arXiv:2008.07782 [hep-ph].

\bibitem{molecular=oset}
  R.~Molina and E.~Oset,
  arXiv:2008.11171 [hep-ph].

\bibitem{molecular=oset2}
  R.~Molina, T.~Branz and E.~Oset,
  Phys.\ Rev.\ D {\bf 82}, 014010 (2010)
  doi:10.1103/PhysRevD.82.014010
  [arXiv:1005.0335 [hep-ph]].

\bibitem{molecular=sundu}
  S.~S.~Agaev, K.~Azizi and H.~Sundu,
  arXiv:2008.13027 [hep-ph].

\bibitem{molecular=Burns}
  T.~J.~Burns and E.~S.~Swanson,
  arXiv:2009.05352 [hep-ph].

\bibitem{molecular=Mutuk}
  H.~Mutuk,
  arXiv:2009.02492 [hep-ph].


\bibitem{tetraquark=Chen}
  H.~X.~Chen, W.~Chen, R.~R.~Dong and N.~Su,
  arXiv:2008.07516 [hep-ph].

\bibitem{tetraquark=Karliner}
  M.~Karliner and J.~L.~Rosner,
  arXiv:2008.05993 [hep-ph].

\bibitem{molecular=Xue}
  Y.~Xue, X.~Jin, H.~Huang and J.~Ping,
  arXiv:2008.09516 [hep-ph].

\bibitem{tetraquark=Zhang}
  J.~R.~Zhang,
  arXiv:2008.07295 [hep-ph].

\bibitem{tetraquark=zhuruilin}
  X.~G.~He, W.~Wang and R.~Zhu,
  arXiv:2008.07145 [hep-ph].

\bibitem{tetraquark=Agaev}
  S.~S.~Agaev, K.~Azizi, B.~Barsbay and H.~Sundu,
  Phys.\ Rev.\ D {\bf 101}, no. 9, 094026 (2020)
  doi:10.1103/PhysRevD.101.094026
  [arXiv:1912.07656 [hep-ph]].

\bibitem{tetraquark=Lu}
  Q.~F.~L\"{u}, D.~Y.~Chen and Y.~B.~Dong,
  arXiv:2008.07340 [hep-ph].

 \bibitem{quark-model-salamenca}
  J.~Vijande, F.~Fernandez and A.~Valcarce,
  J.\ Phys.\ G {\bf 31}, 481 (2005).

\bibitem{quark-modle-our}
  Y.~Yang, C.~Deng, J.~Ping and T.~Goldman,
  Phys.\ Rev.\ D {\bf 80}, 114023 (2009).
  doi:10.1103/PhysRevD.80.114023

\bibitem{3p0}
  X.~Chen, J.~Ping, C.~D.~Roberts and J.~Segovia,
  Phys.\ Rev.\ D {\bf 97}, no. 9, 094016 (2018)
  doi:10.1103/PhysRevD.97.094016
  [arXiv:1712.04457 [nucl-th]].

\bibitem{GEM}
  E.~Hiyama, Y.~Kino and M.~Kamimura,
  Prog.\ Part.\ Nucl.\ Phys.\  {\bf 51}, 223 (2003).
  doi:10.1016/S0146-6410(03)90015-9

\bibitem{RSm}
  E.~Hiyama, M.~Kamimura, A.~Hosaka, H.~Toki and M.~Yahiro,
  Phys.\ Lett.\ B {\bf 633}, 237 (2006)
  doi:10.1016/j.physletb.2005.11.086
  [hep-ph/0507105].

\end{thebibliography}
\end{document}